\documentclass[11pt, reqno]{article}
\pdfoutput=1

\usepackage{graphicx}
\usepackage{jheppub}
\usepackage{amssymb}
\usepackage{amsmath}
\usepackage[usenames,dvipsnames]{xcolor}
\usepackage{epsfig}
\usepackage{dcolumn}
\usepackage{tikz}
\usetikzlibrary{shapes.geometric, arrows}
\usepackage{upgreek}
\usepackage{setspace}
\usepackage{enumitem}
\usepackage{array,multirow,bigdelim,arydshln}
\usepackage{appendix}
\usepackage{xparse}
\usepackage[utf8]{inputenc}
\usepackage{hyperref}
\hypersetup{
	colorlinks,
	urlcolor=Maroon,
	linkcolor=Maroon,
	citecolor=Maroon
}

\usepackage{amsthm}

\theoremstyle{definition}

\usepackage{mathtools}

\usepackage{float}
\restylefloat{table}

\NewDocumentCommand{\binomial}{omm}
 {%
  \genfrac(){0pt}{}{#2}{#3}%
  \IfValueT{#1}{_{\!#1}}%
 }
\NewDocumentCommand{\eulerian}{omm}
 {%
  \genfrac<>{0pt}{}{#2}{#3}%
  \IfValueT{#1}{_{\!#1}}%
 }

\def \s {\sigma}

\usepackage{underscore}

\usepackage{latexsym}
\usepackage{tikz}

\title{Connecting Infinity to Soft Factors}

\author{Freddy Cachazo}\emailAdd{fcachazo@pitp.ca}
\author{and Pablo Leon}\emailAdd{pleon@pitp.ca}

\affiliation{Perimeter Institute for Theoretical Physics, Waterloo, ON N2L 2Y5, Canada}

\abstract{In this note we study tree-level scattering amplitudes of gravitons under a natural deformation which in the large $z$ limit can be interpreted either as a $k$-hard-particle limit or as a $(n-k)$-soft-particle limit. When $k=2$ this becomes the standard BCFW deformation while for $k=3$ it leads to the Risager deformation. The hard- to soft-limit map we define motivates a way of computing the leading order behavior of amplitudes for large $z$ directly from soft limits. We check the proposal by applying the $k=3$ and $k=4$ versions to NMHV and N$^2$MHV gravity amplitudes respectively. The former reproduces in a few lines the result recently obtained by using CHY-like techniques in \cite{BCL}. The N$^2$MHV formula is also remarkably simple and we give support for it using a CHY-like computation. In the $k=2$ case applied to any gravity amplitude, the multiple soft-limit analysis reproduces the correct ${\cal O}(z^{-2})$ behavior while explicitly showing the source of the mysterious cancellation among Feynman diagrams that tames the behavior from the ${\cal O}(z^{n-5})$ of individual Feynman diagrams down to the ${\cal O}(z^{-2})$ of the amplitude.}

\begin{document}
\maketitle
\addtocontents{toc}{\protect\setcounter{tocdepth}{1}}
\def \tr {\nonumber\\}
\def \nn {\nonumber}
\def \la {|}
\def \ra {|}
\def \dd {\Theta}
\def\hset{\texttt{h}}
\def\gset{\texttt{g}}
\def\sset{\texttt{s}}
\def\A {\textsf{A}}
\def\B {\textsf{B}}
\def\C {\textsf{C}}
\def\D {\textsf{D}}
\def\E {\textsf{E}}
\def\F {\textsf{F}}
\def\G {\textsf{G}}
\def\I {\textsf{I}}
\def\J {\textsf{J}}
\def\H {\textsf{H}}
\def \be {\begin{equation}}
\def \ee {\end{equation}}
\def \ba {\begin{eqnarray}}
\def \ea {\end{eqnarray}}
\def \k {\kappa}
\def \h {\hbar}
\def \r {\rho}
\def \l {\lambda}
\def \be {\begin{equation}}
\def \en {\end{equation}}
\def \bes {\begin{eqnarray}}
\def \ens {\end{eqnarray}}
\def \red {\color{Maroon}}
\def \pt {{\rm PT}}
\def \s {\textsf{s}}
\def \t {\textsf{t}}
\def \C {\textsf{C}}
\def \tp {||}
\def \p {x}
\def \x {z}
\def \V {\textsf{V}}
\def \ls {{\rm LS}}
\def \ma {\Upsilon}
\def \SL {{\rm SL}}
\def \GL {{\rm GL}}
\def \w {\omega}
\def \e {\epsilon}
\def \a {\alpha}
\def \g {\gamma}
\def \b {\beta}
\def \ort {\textsf{O}}

\numberwithin{equation}{section}

\section{Introduction}

Tree level scattering amplitudes of massless particles in four dimensions have been the subject of intensive study since Witten's introduction of twistor string theory in 2003 \cite{Witten:2003nn} (see \cite{Elvang:2013cua} for a review). Two of the early developments inspired by twistor string theory were the Cachazo-Svrcek-Witten (CSW) diagrams \cite{Cachazo:2004kj} and the Britto-Cachazo-Feng-Witten recursion relations (BCFW) \cite{Britto:2004ap,Britto:2005fq}. Both techniques were first applied to gluon amplitudes and, while seemingly disconnected, the CSW expansion was shown to follow from a BCFW-like recursion by Risager in \cite{Risager:2005vk}. 

BCFW-like recursion relations are obtained by deforming the kinematic data to build a function of a one-complex dimensional parameter $A(z)$. At tree level, $A(z)$ can only have simple poles for $z\in \mathbb{C}$ with residues determined by factorization properties. When there is no pole at $z=\infty$, the original amplitude is determined using the residue theorem. 

A typical BCFW deformation takes the form
\begin{equation}\label{typicalBCFW}
    \{ \lambda_1,\tilde\lambda_1 \} \!\to\! \{ \lambda_1+ z \lambda_2, \tilde\lambda_1\}  \quad {\rm and}\quad  \{ \lambda_2,\tilde\lambda_2 \} \!\to\! \{ \lambda_2, \tilde\lambda_2-z\tilde\lambda_1\},
\end{equation}
while a Risager deformation is given by 
\begin{equation}\label{typicalRis}
    \{ \lambda_1,\tilde\lambda_1 \} \!\to\! \{ \lambda_1, \tilde\lambda_1 + z \langle 2,3\rangle\tilde\mu \},~ \{ \lambda_2,\tilde\lambda_2 \} \!\to\! \{ \lambda_2, \tilde\lambda_2 + z \langle 3,1\rangle\tilde\mu \}, ~ \{ \lambda_3,\tilde\lambda_3 \} \!\to\! \{ \lambda_3, \tilde\lambda_3 + z \langle 1,2\rangle\tilde\mu \}.
\end{equation}

What makes some BCFW-like recursion relations surprising is the fact that the behavior at infinity can be very different from that of individual Feynman diagrams used in the traditional computation of the amplitude. One of the most striking examples is that of graviton amplitudes under \eqref{typicalBCFW}, where Feynman diagrams can blow up as ${\cal O}(z^{n-5})$ while the amplitude can be shown to vanish as ${\cal O}(z^{-2})$ \cite{Benincasa:2007qj,Arkani-Hamed:2008bsc}. In the gluon case, one can argue that Feynman diagrams vanish as ${\cal O}(z^{-1})$ which is also the correct behavior of the amplitude \cite{Britto:2005fq}.  

Under Risager's deformation \eqref{typicalRis}, the behavior is also unexpected at first. NMHV gluon amplitudes go as ${\cal O}(z^{-4})$ while graviton amplitudes vanish at infinity for small values of $n$  leading to the expectation that perhaps a CSW expansion existed for them \cite{Bjerrum-Bohr:2005xoa}. Unfortunately, NMHV graviton amplitudes stop vanishing at infinity for $n=12$ \cite{Bianchi:2008pu,Benincasa:2007qj}. In fact, NMHV graviton amplitudes behave as ${\cal O}(z^{n-12})$ for large $z$. This shows that for $n>11$, graviton amplitudes receive contributions from the residue of a pole of order $n-11$ at infinity\footnote{The order of the pole is computed for the function $A(z)/z$, which is the one used in BCFW-like constructions.}. Such a residue for $n=12$ was first computed by Conde and Rajabi \cite{Conde:2012ik} and more recently in \cite{BCL} by Belayneh and the authors (BCL).  

The BCL formula provides a simple analytic form for the leading order in $1/z$ behavior of the amplitude. In \cite{BCL}, it was also noted that the leading order formula can be written in a way that suggests a direct connection to multi-soft limits \cite{Klose:2015xoa,Zlotnikov:2017ahq,Saha:2017yqi,AtulBhatkar:2018kfi,Chakrabarti:2017ltl}. More explicitly, the leading order has the schematic form
\be
A_n^{\rm NMHV}(z)=\frac{1}{z^4}\sum_{4\leq a<b\leq n}\!\!\!A_5(1^-,2^-,3^-,a^+,b^+)(z)\prod_{\substack{c=4 \\ c\neq a,b}}^n{\cal S}_c(z)+\ldots
\ee 
where ${\cal S}_c(z)$ are Weinberg's soft graviton factors \cite{Weinberg:1964ew}. The precise formula is given in \eqref{loBCL}.

In this note we explore the connection between the behavior of graviton scattering amplitudes for large $z$ under various deformations and multi-soft limits (for related work see \cite{Arkani-Hamed:2008bsc,Mason:2009afn,Cohen:2010mi}). The connection stems from a simple use of scale and little group transformations. In section \ref{sec:map} we derive the explicit map that connects the two problems. 

The deformations we consider unify the standard BCFW  and Risager's deformations as the first two cases of a family of deformations characterized by $k$, the number of deformed particles. Our family of deformations can be interpreted as a multi-hard limit of $k$ particles and also, using the map in section \ref{sec:map}, as a multi-soft limit of $n-k$ particles.   

In section 3, we apply the map to the Risager deformation of NMHV graviton amplitudes. We obtain the correct large $z$ behavior matching the BCL analytic formula while explaining why Weinberg's soft factors showed unexpectedly in the BCL formula. 

In these applications the multi-soft limits are not given by known formulas \cite{Chakrabarti:2017ltl} since the orders governed by soft theorems vanish. Here we propose a prescription for how to compute them and use the perfect agreement with the BCL formula as a strong consistency check.     

In section \ref{sec:N2MHV}, we apply the map developed in section \ref{sec:map} to the $k=4$ case and use it to study N$^2$MHV graviton amplitudes. A multi-soft limit analysis analogous to that of the NMHV case leads to a prediction for the leading order form in $1/z$ of the amplitude. More explicitly,
\be
A^{{\rm N}^2{\rm MHV}}_n(z)=\frac{1}{z^6}\!\sum_{5\leq a<b\leq n}\!\!\!A_6(1^-,2^-,3^-,4^-,a^+,b^+)(z)\prod_{\substack{c=5 \\ c\neq a,b}}^n{\cal S}_c(z)+\ldots
\ee 
The detailed formula is presented in \eqref{n2MHV}.

We reproduce this formula by using a combination of scattering equations \cite{Cachazo:2013hca}, Witten-RSV equations \cite{Witten:2003nn,Roiban:2004yf} and the Cachazo-Skinner-Mason formula \cite{Cachazo:2012kg,Cachazo:2012pz} for graviton amplitudes following the same steps done in \cite{BCL} for NMHV amplitudes. The N$^2$MHV scattering equations have a much more intricate structure than the NMHV ones.  

In section \ref{sec:bcfw}, we apply the map to BCFW deformations and re-derive the enigmatic large $z$ behavior of graviton amplitudes from Weinberg's soft factors. A simple counting argument reveals that the worse behavior of individual Feynman diagrams is ${\cal O}(z^{n-5})$ as $z\to \infty$. However, it is known that graviton amplitudes vanish as ${\cal O}(z^{-2})$. The source of this huge cancellation is not explained by any of the approaches that prove the behavior of the amplitude. One of the reasons is that the known proofs do not use Feynman diagrams \cite{Benincasa:2007qj,Arkani-Hamed:2008bsc}. While our computation does not directly use Feynman diagrams either, the soft factors are very close to them and the cancellation we see in each factor clearly shows how it happens among Feynman diagrams. 

In section \ref{sec:disc} we discuss extensions of this work. In particular, given that Weinberg's soft factor \cite{Weinberg:1964ew} controls the leading order in the soft limit and  sub- and sub-sub-leading terms are also known \cite{Cachazo:2014fwa}, we discuss the possibility of using these subleading terms to obtain higher order expressions in the large $z$ expansion. 

\section{Deformations and The Hard-Soft Map}\label{sec:map}

In this section we introduce the deformations of interest and the interpretation  of their large $z$ limit as $k$ multiple-hard or as $n-k$ multiple-soft limits (for related work see \cite{Arkani-Hamed:2008bsc,Mason:2009afn,Cohen:2010mi}). 

We start partitioning the set of $n$ particles into deformed, ${\cal H}$, and undeformed, ${\cal S}$, particles. Let $k:=|{\cal H}|$ and so $n-k=|{\cal S}|$. 

The deformed kinematic data is defined as
\begin{equation}\label{defor1}
    \{ \lambda_i(z),\tilde\lambda_i(z) \} := \{ \lambda_i+z \mu_i, \tilde\lambda_i + z \tilde\mu_i\} \quad {\rm for} \quad i\in {\cal H}.
\end{equation}
Without loss of generality, let ${\cal H}=\{1,2,\ldots ,k\}$. Momentum conservation requires
\begin{equation}
    \sum_{i=1}^k \lambda_i(z)\tilde\lambda_i(z) + \sum_{a=k+1}^n \lambda_a\tilde\lambda_a = 0
\end{equation}
for all $z$. This is a set of four degree-two polynomials in $z$ and hence each coefficient must vanish by itself. The resulting equations are
\begin{align}
    & \label{zsquare} \sum_{i=1}^k \mu_i\tilde\mu_i = 0, \\
    & \label{zone}\sum_{i=1}^k \left( \lambda_i\tilde\mu_i + \mu_i\tilde\lambda_i \right) = 0, \\
    & \label{zero}\sum_{i=1}^k \lambda_i\tilde\lambda_i + \sum_{a=k+1}^n \lambda_a\tilde\lambda_a = 0.
\end{align}
Note that \eqref{zsquare} is nothing but the momemtun conservation conditions for a $h$-particle amplitude while \eqref{zero} is that for a $n$-particle amplitude. Although somewhat unfamiliar, \eqref{zone} can also be interpreted as momentum conservation of a $2k$-particle amplitude. 

Now consider a generic $n$-particle amplitude with particle helicities $h_a$ and apply the deformation \eqref{defor1} to define a function of $z$,
\be 
F(z) := A(\ldots , \{ \lambda_i(z),\tilde\lambda_i(z), h_i \},\ldots,  \{ \lambda_a,\tilde\lambda_a, h_a \},\ldots ).
\ee 

While we are interested in the large $z$ behavior of $F(z)$, we will not attempt a direct approach and instead we map it to a different problem. 

For simplicity, let us assume that all particles have the same spin, i.e. $s:=|h_i| = |h_a|$ and recall that under little group and scale transformations, a general amplitude transforms as 
\begin{align}\label{littleG} 
\nonumber A(\{ t_a\lambda_a,t_a^{-1}\tilde\lambda_a, h_a \}) & =  t_a^{-2h_a}A(\{ \lambda_a,\tilde\lambda_a, h_a \}), \\
A(\{ \rho^{1/2}\lambda_a,\rho^{1/2}\tilde\lambda_a, h_a \}) & = \rho^{(n-3)(s-2)+s}A(\{ \lambda_a,\tilde\lambda_a, h_a \}),
\end{align}
respectively.

Using the scale transformation in \eqref{littleG} on $F(z)$ one can show that 
\be\label{firstF}
F(z) = z^{(n-3)(s-2)+s}A(\ldots , \{ z^{-1/2}\lambda_i(z),z^{-1/2}\tilde\lambda_i(z), h_i \},\ldots,  \{ z^{-1/2}\lambda_a,z^{-1/2}\tilde\lambda_a, h_a \},\ldots ).
\ee 

As a last step, we assume that the deformation only happens in either the holomorphic or anti-holomorphic spinors, i.e., either $\mu_i =0$ or $\tilde\mu_i=0$ in \eqref{defor1}. In the former (latter) case, we call the deformation holomorphic (anti-holomorphic). Let us rewrite the spinors in \eqref{firstF} according to their deformation as
\begin{align}
\{ z^{-1/2}\lambda_i(z),z^{-1/2}\tilde\lambda_i,h_i\} & =  \{ z^{1/2}\left(z^{-1}\lambda_i(z)\right),z^{-1/2}\tilde\lambda_i,h_i\} = \{ z^{1/2}\left(\mu_i + z^{-1}\lambda_i\right),z^{-1/2}\tilde\lambda_i,h_i\}, \\
\{ z^{-1/2}\lambda_i,z^{-1/2}\tilde\lambda_i(z),h_i\} & =  \{ z^{-1/2}\lambda_i,z^{1/2}\left( z^{-1}\tilde\lambda_i(z)\right),h_i\}=\{ z^{-1/2}\lambda_i,z^{1/2}\left( \tilde\mu_i + z^{-1}\tilde\lambda_i\right),h_i\}.
\end{align}
These expressions motivate the introduction of the following notation $\mu_i(z):=\mu_i + z^{-1}\lambda_i$ and $\tilde\mu_i(z):=\tilde\mu_i + z^{-1}\tilde\lambda_i$. 

Using a little group transformation \eqref{littleG} on the expression for $F(z)$ given in \eqref{firstF} one can remove the factors of $z^{1/2}$ and $z^{-1/2}$ from the particles in ${\cal H}$  to get
\begin{align}\label{secF}
\nonumber
F(z) & = z^{(n-3)(s-2)+s}\left(\prod_{i\in {\cal H}}z^{-\epsilon_i h_i}\right)\times \\
& A(\ldots , \{ \lambda_i,\tilde\mu_i(z), h_i \},\ldots ,\{ \mu_j(z),\tilde\lambda_j, h_j \}\ldots ,  \{ z^{-1/2}\lambda_a,z^{-1/2}\tilde\lambda_a, h_a \},\ldots ).
\end{align}

Here $\epsilon_i =1$ if a holomorphic deformation is selected and $\epsilon_i =-1$ for an anti-holomorphic one.

This completes the map. Now we can see that in order to study the large $z$ behavior of $F(z)$, which can be thought of as the hard limit of particles in ${\cal H}$, it suffices to study that of 
\be 
G(z) :=A(\ldots , \{ \lambda_i,\tilde\mu_i(z), h_i \},\ldots ,\{ \mu_j(z),\tilde\lambda_j, h_j \}\ldots ,  \{ z^{-1/2}\lambda_a,z^{-1/2}\tilde\lambda_a, h_a \},\ldots ) 
\ee 
which is nothing but the soft limit of particles in the set ${\cal S}$, while keeping particles in the set ${\cal H}$ as the hard ones but with small deformations that ensure momentum conservation during the limit. 

\section{NMHV Risager Deformation}

In \cite{BCL}, the leading order behavior of NMHV graviton amplitudes under the Risager deformation \eqref{typicalRis} was computed and an explicit analytic formula was presented for all $n$. In this section we show how the same formula can be reproduced by an argument that uses the map introduced in section \ref{sec:map} and Weinberg's soft factors.

Consider the graviton amplitude $A(1^-,2^-,3^-,4^+,5^+,\ldots ,n^+)$ under the Risager deformation \eqref{typicalRis}, which coincides with the $k=3$ all anti-holomorphic deformation of the previous section, and apply the map to get
\be  
F(z) = z^{-4}G(z)
\ee 
with 
\be\label{nmhvG} 
G(z) = A(\{ \lambda_1,\langle 2,3\rangle \tilde\mu + \tilde\lambda_1/z,-2\},\{ \lambda_2,\langle 3,1\rangle \tilde\mu+ \tilde\lambda_2/z,-2\},\{ \lambda_3,\langle 1,2\rangle \tilde\mu+ \tilde\lambda_3/z,-2\}, \ldots ) 
\ee 
and where ellipses represent particles $4,5,\ldots ,n$ with kinematic data
$\{ z^{-1/2}\lambda_a,z^{-1/2}\tilde\lambda_a,+2\}$. 

The large $z$ limit of $G(z)$ is nothing but a multi-soft limit where particles $4,5,\ldots ,n$ are soft. 

It is know that using Weinberg's soft theorems the leading order of multiple simultaneous soft limits can be computed by applying sequential soft limits on the ``hard'' amplitude if the latter is not zero. In our case this condition is not satisfied since $A(1^-,2^-,3^-)=0$ in Einstein gravity. This means that the procedure has to be modified. Here we propose that the simultaneous soft limit of particles $4,5,\ldots ,n$ is computed by taking as many consecutive soft limits as possible to get a non-vanishing ``hard" amplitude and summing over all ways to achieve this. In our case, such a non-vanishing amplitude has the form $A(1^-,2^-,3^-,i^+,j^+)$ with $i,j\in \{ 4,5,\ldots ,m \}$. Once the sum over all possible choices of $i,j$ is made, the result is evaluated on the $z$ dependent kinematics and the leading order is extracted. 

Let us now explicitly carry out the procedure. The leading behavior under soft limits is governed by Weinberg's soft factor \cite{Weinberg:1964ew}. Assuming particle $a$ is soft, with $h_a=2$ and that the ``hard" particles are $1,2,3,4,5$, we have
\be\label{weinb} 
{\cal S}_a(z) = \sum_{\substack{b=1 \\ b\neq a}}^5 \frac{[a,b](z)}{\langle a,b \rangle(z)}\frac{\langle b,x \rangle\langle b,y \rangle(z)}{\langle a,x \rangle(z)\langle a,y \rangle(z)}
\ee 
where $\lambda_x$ and $\lambda_y$ are reference spinors. The choice of such spinors is irrelevant provided the particles involved in \eqref{weinb} satisfy momentum conservation. If $h_a=-2$ then one simply performs the transformation $[i,j]\leftrightarrow \langle i,j \rangle$.

Plugging in the kinematic data in $G(z)$ \eqref{nmhvG} and using $(i,j,k)$ to denote an ordered set which is a cyclic permutation of $(1,2,3)$, one finds
\be 
{\cal S}_a(z) = z\sum_{i=1}^3 \frac{(\langle j,k \rangle[a,\tilde\mu]+[a,i]/z )}{\langle a,i \rangle}\frac{\langle i,x \rangle\langle i,y \rangle}{\langle a,x \rangle\langle a,y \rangle}+\sum_{b=4}^5 \frac{[a,b]}{\langle a,b \rangle}\frac{\langle b,x \rangle\langle b,y \rangle}{\langle a,x \rangle\langle a,y \rangle}.
\ee 
Keeping only the leading order we arrive at,
\be
{\cal S}_a(z) = z\sum_{i=1}^3 \frac{\langle j,k \rangle[a,\tilde\mu] }{\langle a,i \rangle}\frac{\langle i,x \rangle\langle i,y \rangle}{\langle a,x \rangle\langle a,y \rangle}+{\cal O}(z^0).
\ee 
Note that changing the reference spinors $\lambda_x$ and $\lambda_y$ only affects the ${\cal O}(z^0)$ terms. The soft factors can be further simplified by choosing $x=1$ and $y=2$,
\be\label{softLO} 
{\cal S}_a(z) = z [a,\tilde\mu]\frac{\langle 1,2 \rangle\langle 1,3 \rangle\langle 2,3 \rangle}{\langle 1,a \rangle\langle 2,a \rangle\langle 3,a \rangle}+{\cal O}(z^0).
\ee 

Using Weinberg's theorem consecutively and summing over all ways of singling out $a,b$ we arrive at
\be 
G(z) =\!\! \sum_{4\leq a<b\leq n}\!\!\!\!\!A(1^-,2^-,3^-,a^+,b^+)(z)\prod_{\substack{c=4 \\ c\neq a,b}}^n{\cal S}_c(z)+\ldots
\ee 

Let us now compute $A(1^-,2^-,3^-,a^+,b^+)(z)$ and expand it around $z=\infty$. The explicit amplitude, before adding the $z$ dependence, can easily be written down using Hodges' formula \cite{Hodges:2009hk}, 
\be
A(1^-,2^-,3^-,a^+,b^+) = [ a,b]^8\frac{\frac{\langle 1,a\rangle}{[1,a]}\frac{\langle 2,b\rangle}{[2,b]}-\frac{\langle 2,a\rangle}{[2,a]}\frac{\langle 1,b\rangle}{[1,b]}}{[1,2][2,3][3,1][3,a][a,b][b,3]}.
\ee 
Adding in the $z$ dependence inherited from $G(z)$ and expanding one finds
\be\label{fiveZ}
A(1^-,2^-,3^-,a^+,b^+)(z) =z^{-3}\frac{[ a,b]^7\langle a,b\rangle}{\langle 1,2\rangle\langle 1,3\rangle\langle 2,3\rangle[\tilde\mu,a]^2[\tilde\mu,b]^2\langle 1|P|\tilde\mu ]\langle 2|P|\tilde\mu ]\langle 3|P|\tilde\mu ]} + {\cal O}(z^{-4}).
\ee 
Here we used that e.g. $[1,2](z) = \langle 3|P|\tilde\mu ]/z$ with $P:=k_1+k_2+k_3$, $\langle 1,a\rangle(z)=z^{-1/2}\langle 1,a\rangle$, and $[3,a](z) =\langle 1,2\rangle [\tilde\mu,a]$. Recall that all these statements are to leading order in $1/z$.
Plugging \eqref{fiveZ} into $G(z)$ and then the result into $F(z)$ we arrive at the final formula for the leading order of the NMHV amplitude under a Risager deformation
\be\label{loBCL}
\left. F(z)\right|_{\rm L.O.} = z^{n-12}\!\!\!\! \sum_{4\leq a <b\leq n}\!\! \frac{[ a,b]^7\langle a,b\rangle(\langle 1,2\rangle\langle 1,3\rangle\langle 2,3\rangle)^{n-6}}{[\tilde\mu,a]^2[\tilde\mu,b]^2\langle 1|P|\tilde\mu ]\langle 2|P|\tilde\mu ]\langle 3|P|\tilde\mu ]}\prod_{\substack{c=4 \\ c\neq a,b}}^n\frac{[c,\tilde\mu]}{\langle 1,c \rangle\langle 2,c \rangle\langle 3,c \rangle}.
\ee 

This is the formula recently computed in \cite{BCL} using the Cachazo-Skinner-Mason formulation of gravity amplitudes \cite{Cachazo:2012kg,Cachazo:2012pz}, which agrees for $n=12$ with the Conde-Rajabi formula for the residue at $z=\infty$ of the Risager deformed amplitude \cite{Conde:2012ik}.

\section{N$^2$MHV Generalized Risager Deformation}\label{sec:N2MHV}

Having successfully reproduced the NMHV leading order behavior, it is natural to consider the next sector. The deformation now involves four particles, i.e., $k=4$. Using the general formula \eqref{defor1} and the conditions \eqref{zsquare}, \eqref{zone}, \eqref{zero} with ${\cal H}=\{ 1,2,3,4\}$, $\mu_i=0$ and $\tilde\mu_i\neq 0$ gives
\be\label{grd}
\tilde\lambda_i(z):=\tilde\lambda_i + z \tilde\mu_i \quad {\rm for} \quad i\in {\cal H}\quad {\rm and} \quad \sum_{i=1}^4 \lambda_i\tilde\mu_i  = 0.
\ee

The most general choice of $\tilde\mu_i$'s gives rise to general four-particle kinematics $\{ \lambda_i, \tilde\mu_i\}$ subject to momentum conservation \eqref{grd}. Note that this is different from Risager's deformation \cite{Risager:2005vk} for $k=4$ where $\tilde\mu_i =\alpha_i\tilde\mu$. In other words, in Risager's case, $[\tilde\mu_i,\tilde\mu_j]=0$ for all $i,j\in\{1,2,3,4\}$. This means that from the ``hard" point of view, Risager's deformation is very singular and this is why we do not use it.   

Repeating the same analysis as in the previous section and making the same assumption that the leading order terms are computed by carrying out the soft limits of all but two of the positive helicity gravitons leads to 
\be 
G(z) = \!\! \sum_{5\leq a<b\leq n}\!\!\!\!\!A(1^-,2^-,3^-,4^-,a^+,b^+)(z)\prod_{\substack{c=5 \\ c\neq a,b}}^n{\cal S}_c(z)+\ldots
\ee 
while using the map gives
\be 
F(z)=z^{-6}G(z).
\ee 
Without loss of generality, let us consider the six-graviton amplitude with $a=5$ and $b=6$, 
\be 
A(1^-,2^-,3^-,4^-,5^+,6^+)=[5,6]^8\frac{\begin{vmatrix}
\frac{\langle 1,4\rangle}{[1,4]} & \frac{\langle 1,5\rangle}{[1,5]} & \frac{\langle 1,6\rangle}{[1,6]}\\
\frac{\langle 2,4\rangle}{[2,4]} & \frac{\langle 2,5\rangle}{[2,5]} & \frac{\langle 2,6\rangle}{[2,6]}\\
\frac{\langle 3,4\rangle}{[3,4]} & \frac{\langle 3,5\rangle}{[3,5]} & \frac{\langle 3,6\rangle}{[3,6]}
\end{vmatrix}}{[ 1,2] [1,3] [2,3] [4,5] [4,6] [5,6]} . 
\ee 
Applying the $z$ deformation, the six-graviton amplitude to leading order in $1/z$ becomes 
\be 
A(1^-,2^-,3^-,4^-,5^+,6^+)(z)=\frac{1}{z^6}[5,6]^7\frac{\begin{vmatrix}
\frac{\langle 1,4\rangle}{[\tilde\mu_1,\tilde\mu_4]} & \frac{\langle 1,5\rangle}{[\tilde\mu_1,5]} & \frac{\langle 1,6\rangle}{[\tilde\mu_1,6]}\\
\frac{\langle 2,4\rangle}{[\tilde\mu_2,\tilde\mu_4]} & \frac{\langle 2,5\rangle}{[\tilde\mu_2,5]} & \frac{\langle 2,6\rangle}{[\tilde\mu_2,6]}\\
\frac{\langle 3,4\rangle}{[\tilde\mu_3,\tilde\mu_4]} & \frac{\langle 3,5\rangle}{[\tilde\mu_3,5]} & \frac{\langle 3,6\rangle}{[\tilde\mu_3,6]}
\end{vmatrix}}{[ \tilde\mu_1,\tilde\mu_2] [\tilde\mu_1,\tilde\mu_3] [\tilde\mu_2,\tilde\mu_3] [\tilde\mu_4,5] [\tilde\mu_4,6] } + \ldots 
\ee 

Let us introduce a special notation for the leading order  
\be
B(1^-,2^-,3^-,4^-,5^+,6^+):=[5,6]^7\frac{\begin{vmatrix}
\frac{\langle 1,4\rangle}{[\tilde\mu_1,\tilde\mu_4]} & \frac{\langle 1,5\rangle}{[\tilde\mu_1,5]} & \frac{\langle 1,6\rangle}{[\tilde\mu_1,6]}\\
\frac{\langle 2,4\rangle}{[\tilde\mu_2,\tilde\mu_4]} & \frac{\langle 2,5\rangle}{[\tilde\mu_2,5]} & \frac{\langle 2,6\rangle}{[\tilde\mu_2,6]}\\
\frac{\langle 3,4\rangle}{[\tilde\mu_3,\tilde\mu_4]} & \frac{\langle 3,5\rangle}{[\tilde\mu_3,5]} & \frac{\langle 3,6\rangle}{[\tilde\mu_3,6]}
\end{vmatrix}}{[ \tilde\mu_1,\tilde\mu_2] [\tilde\mu_1,\tilde\mu_3] [\tilde\mu_2,\tilde\mu_3] [\tilde\mu_4,5] [\tilde\mu_4,6] } .
\ee

Combining all the pieces, 

\be  \label{n2MHV}
F(z) = z^{n-18}\!\! \sum_{5\leq a<b\leq n}\!\!\!\!\! B(1^-,2^-,3^-,4^-,a^+,b^+)\prod_{\substack{c=5 \\ c\neq a,b}}^n\sum_{i=1}^4\frac{[\tilde\mu_i,c]\langle i,a\rangle\langle i,b\rangle}{\langle i,c\rangle\langle c,a\rangle\langle c,b\rangle}+ {\cal O}(z^{n-19}).
\ee 

Eq. \ref{n2MHV} is our proposal for the leading order as $z\to \infty$ of a $n$-graviton amplitude with helicities $1-,2-,3-,4-,5+,\ldots ,n+$ using the $k=4$ deformation. As explained in the previous section, the validity of this formula relies on that of our prescription for computing the multi-soft limit when the hard amplitude vanishes. Next, we follow the technique developed in \cite{BCL} to provide independent evidence for \ref{n2MHV}.

\subsection{Computation Using Witten-RSV Equations and the CSM Formula}

In this section, we shall use the equations in terms of rationals maps proposed by Witten in \cite{Witten:2003nn} and developed by Roiban, Spradlin, and Volovich (RSV) in \cite{Roiban:2004yf}, whose solutions correspond to the ones in the $\text{N}^{k-2}\text{MHV}$ sector. These are
\begin{align} \label{wrv}
   & \lambda_a = t_a \lambda(u_a) \quad {\rm for}\quad a\in \{ 1,2,\ldots ,n\}, \\
    & \sum_{a=1}^n t_a\tilde\lambda_a u_a^m = 0 \quad {\rm for}\quad m\in \{ 0,1,\ldots ,d\},
\end{align} 
with $u$ denoting an inhomogenous coordinate on $\mathbb{CP}^1$ and $\lambda(z)\in \mathbb{C}^2$ a rational map of degree $d=k-1$ onto $\mathbb{CP}^1$ defined by, 
\be
    \lambda(u)  :=  \rho_0+\rho_1 u+\ldots +\rho_d u^d,
\ee
where $\rho_m$ are spinors. The number of solutions for each $d$ is given by $E(n-3,d-1)$, where $E(a,b)$ is an Eulerian number and counts the number of permutations of $a$ elements with $b$ descends.

Based on these equations, a formula for $\text{N}^{k-2}\text{MHV}$ amplitudes in $\mathcal{N}=8$ supergravity was presented by Cachazo, Skinner in \cite{Cachazo:2012kg} and with Mason in \cite{Cachazo:2012pz}. The CSM formula can be written as
\be 
A_n^{(k=d+1)} = \int d \mathcal{M}_{n,d} \>\> |\Phi|'|\tilde{\Phi}|' \prod_{a=1}^n \delta^2(\lambda_a - t_a \sum_{m=0}^d \rho_m u_a^m) \>\> \prod_{m=0}^d \delta^2(\sum_{a=1}^n t_a u_a^m \tilde \lambda_a)\delta^8(\sum_{a=1}^n t_a u_a^m \eta_a),
\ee
where $\eta_a$ is the corresponding eight-component Grassmann variable and the measure is given by
\be 
d \mathcal{M}_{n,d} = \frac{1}{\text{Vol}(GL(2, \mathbb{C}))}\prod_{m=0}^d d^{2}\rho_m \prod_{a=1}^n \frac{dt_a \>\> du_a}{t^3_a}.
\ee
The $\text{Vol}(GL(2, \mathbb{C}))$ is there because of the $GL(2, \mathbb{C})$ redundancy of the Witten-RSV equations that can be fixed by choosing the values of, e.g., $\{ u_1,u_2,u_3,t_1\}$. Here $|\Phi|'$ and $|\tilde{\Phi}|'$ are defined as follows: Let $\Phi$ be the symmetric $n\times n$ matrix with elements 
\bes 
    \Phi_{ij} &=& \frac{\langle i ~ j \rangle}{u_i-u_j}\frac{1}{t_it_j} \quad \mbox{for} \quad i\neq j, \\
    \Phi_{ii} &=& - \sum_{j\neq i} \Phi_{ij}\prod_{r=0}^{n-d-2} \left(\frac{u_j-v_{r}}{u_i-v_r} \right) \frac{\prod_{k \neq i} (u_i-u_{k})}{\prod_{l \neq j} (u_j-u_{k})}
\ens
where $v_{r}$ are reference points on $\mathbb{CP}^1$ and $\tilde{\Phi}$ the $n\times n$ matrix with elements
\bes
    \tilde{\Phi}_{ij} &=&  \frac{[ i ~ j]}{u_i-u_j}t_it_j \quad \mbox{for} \quad i\neq j, \\
    \tilde{\Phi}_{ii} &=& - \sum_{j\neq i} \tilde{\Phi}_{ij}\prod_{a=0}^{d} \frac{u_j-v_a}{u_i-v_a}.
\ens
Then $|\Phi|'$ is defined as
\be
    |\Phi|' := \frac{\det(\Phi_{red})}{|r_1...r_d||c_1...c_d|},
\ee
where $\Phi_{red}$ is a non-singular matrix obtained by removing any $n-d$ rows and columns of $\Phi$. The labels $r_1...r_d$ and $c_1...c_d$ correspond to the rows and columns that remain in $\Phi_{red}$, respectively.  
Similarly, we can define a non-singular matrix, $\tilde{\Phi}_{red}$, removing any $d+2$ rows and columns of $\tilde{\Phi}$ to define $|\tilde{\Phi}|'$ as 

\be
    |\tilde{\Phi}|' := \frac{\det(\tilde \Phi_{red})}{|\tilde{r}_1...\tilde{r}_{d+2}||\tilde{c}_1...\tilde{c}_{d+2}|}.
\ee
In this case, $\tilde{r}_1...\tilde{r}_d$ and $\tilde{c}_1...\tilde{c}_d$ are the labels of the deleted rows and columns of $\tilde{\Phi}$, respectively. 

From now on we will consider only the case $d=3$ corresponding to the N$^2$MHV sector and use the ${\rm GL}(2,\mathbb{C})$ gauge fixing $u_1=0, u_2=\infty$, $u_3=1$ and $t_1=1$. We can solve the Witten-RSV equations starting with \eqref{wrv} and $a\in\{ 1,2,3,4\}$ which give linear equations for all the $\rho$'s,
\begin{eqnarray}
    \rho_0 &=& \lambda_1, \\
    \rho_1 &=& -\lambda_1 \frac{1+u_4}{u_4} + \lambda_2 \frac{u_4}{t_2} -\lambda_3 \frac{u_4}{t_3 (1-u_4)} + \lambda_4 \frac{1}{t_4u_4(1-u_4)}, \\
    \rho_2 &=& \lambda_1 \frac{1}{u_4} - \lambda_2 \frac{1+u_4}{t_2} +\lambda_3 \frac{1}{t_3 (1-u_4)} - \lambda_4 \frac{1}{t_4u_4(1-u_4)}, \\
    \rho_3 &=& \frac{\lambda_2}{t_2}.
\end{eqnarray}

As in the case of the NMHV amplitudes, the solutions to the Witten-RSV equations for $k = 4$ also split into different types, although in a much more complicated way. The simplest type of solutions is characterized by a partition of ${\cal S}= \{5,\ldots ,n\}$ into two sets, one of which contains two elements. Let $J:=\{a,b\}\subset \{5,\ldots ,n\}$ and $\Bar{J}:= \{5,\ldots,n\}\setminus J $ be the two sets. To leading order in $1/z$, for $i\in J$ one has that $u_i$ is identical to the value it takes in the $\overline{\rm MHV}$ solution of the whole deformed system while for $i\in \Bar{J}$, $u_i$ takes the value in the MHV solution. More explicitly,  
\begin{equation}\label{uJsec}
    u_i = \begin{cases} \displaystyle \frac{[2,3](z)[1,i](z)}{[1,3](z)[2,i](z)} = \frac{[\tilde{\mu}_2,\tilde{\mu}_3][\tilde{\mu}_1,i]}{[\tilde{\mu}_1,\tilde{\mu}_3][\tilde{\mu}_2,i]}+ {\cal O}(1/z), 
     & \text{if $i\in J$}
     \\  \\
    \displaystyle \frac{\langle 2,3 \rangle \langle 1,i \rangle}{\langle 1,3 \rangle \langle 2,i \rangle}+ {\cal O}(1/z),  & \text{if $i\in \Bar{J}$},
    \end{cases}
\end{equation}
and, as expected from deformation (\ref{grd}),
\be
 u_4 = \frac{\langle 2,3\rangle \langle 1,4\rangle}{\langle 1,3\rangle \langle 2,4\rangle} +  {\cal O}(1/z).  
\ee
Notice that $u_4$ can also be written as 

\be \label{u4sol}
 u_4 = \frac{[2,3](z)[1,4](z)}{[1,3](z)[2,4](z)} = \frac{[\tilde{\mu}_2,\tilde{\mu}_3][\tilde{\mu}_1,4]}{[\tilde{\mu}_1,\tilde{\mu}_3][\tilde{\mu}_2,4]} +  {\cal O}(1/z).  
\ee
On the other hand

\begin{equation}
t_i = \begin{cases}\displaystyle
			- z\frac{[\tilde{\mu}_1,\tilde{\mu}_4][\tilde{\mu}_1,\tilde{\mu}_3][i,\tilde{\mu}_2]^2\langle i| a+b |\tilde{\mu}_1]}{[\tilde{\mu}_1,\tilde{\mu}_2]^2[i,\tilde{\mu}_3][i,\tilde{\mu}_4]\langle a,b \rangle [a,b]} + {\cal O}(z^0), & \text{if $i\in J$}\\  \\

   \displaystyle
   -\frac{\langle 2, i\rangle^3}{\langle 1, 2\rangle}\prod_{l \in J} \frac{\langle 1, 3\rangle\langle 1, 4\rangle[l, \tilde{\mu}_1]}{\langle 1, 3\rangle\langle 2, 3\rangle \langle 4, i\rangle [l, \tilde{\mu}_3]+\langle 2, 4\rangle\langle 1, 4\rangle \langle 3, i\rangle [l, \tilde{\mu}_4]}+ {\cal O}(1/z), & \text{if $i\in \bar{J}$}

            \\  \\
\displaystyle
            -(-1)^{\delta_{2,i}}\!\!\sum_{\substack{p,q=2 \\ p,q\neq i}}^4\left(r_{i:p,q}\frac{[\tilde{\mu}_1,\tilde{\mu}_p]}{[\tilde{\mu}_i,\tilde{\mu}_p]}\frac{[\tilde{\mu}_i,\tilde{\mu}_q]^2}{[\tilde{\mu}_1,\tilde{\mu}_q]^2}\right) \prod_{l \in J} \frac{[l,\tilde{\mu}_1]}{[l,\tilde{\mu}_i]} + {\cal O}(1/z), & \text{if $i\in \{ 2,3,4 \}$,}
		 \end{cases}   
\end{equation}
where 
\be 
r_{i:p,q} := ((\delta_{i,2}+\delta_{i,3})\delta_{p,4}+\delta_{i,4}\delta_{p,3})((\delta_{i,3}+\delta_{i,4})\delta_{q,2}+\delta_{i,2}\delta_{q,3}).
\ee

The analysis of the cases $n$ = 7, 8, and 9 reveals that these solutions are responsible for the leading order behavior of the amplitude in the large $z$ limit. Assuming that this is true for all $n$, we can use  these expressions, together with the generalized Risager deformation, to get

\begin{eqnarray}
    |\Phi|'&=& -\frac{([\tilde{\mu}_1,\tilde{\mu}_2][\tilde{\mu}_1,\tilde{\mu}_3])^5([\tilde{\mu}_2,\tilde{\mu}_4][\tilde{\mu}_3,\tilde{\mu}_4])^2\prod_{i=2}^4[\tilde{\mu}_i,a]^2[\tilde{\mu}_i,b]^2}{z^2([\tilde{\mu}_1,a][\tilde{\mu}_1,b])^4[a,b]^6[\tilde{\mu}_2,\tilde{\mu}_3]^7[\tilde{\mu}_1,\tilde{\mu}_4]^4} , \nonumber \\ && \hspace{5cm} \times B(1^-,2^-,3^-,4^-,a^+,b^+) + {\cal O}(1/z^3),\\
    |\tilde{\Phi}|' &=& \frac{z^{n-4}[\tilde{\mu}_1,\tilde{\mu}_3]^{13}[\tilde{\mu}_2,\tilde{\mu}_4]^{n}([\tilde{\mu}_2,a][\tilde{\mu}_2,b])^6 \prod_{i\in \Bar{J}}(\langle 2 ,i \rangle t_i)^2}{[\tilde{\mu}_2,\tilde{\mu}_3]^7[\tilde{\mu}_1,\tilde{\mu}_4]^{n-6}[\tilde{\mu}_1,\tilde{\mu}_2]^{11}[\tilde{\mu}_3,\tilde{\mu}_4]^2[a,b]^2\prod_{i=3}^4[\tilde{\mu}_i,a]^2[\tilde{\mu}_i,b]^2} \nonumber \\ && \hspace{5cm} \times \prod_{\substack{c=5 \\ c\neq a,b}}^n\sum_{i=1}^4\frac{[\tilde\mu_i,c]\langle i,a\rangle\langle i,b\rangle}{\langle i,c\rangle\langle c,a\rangle\langle c,b\rangle}  + {\cal O}(z^{n-5}),
\end{eqnarray}
from which the core six-point amplitude and the soft factors of (\ref{n2MHV}) are already manifest. Finally, we can plug it all into the CSM formula, and after a somewhat tedious computation, one obtains \footnote{The measure and the Jacobian associated with the delta functions in the CSM formula contribute with a power of $z^{-6}$ each.}

\begin{eqnarray}
    F^{(a,b)}(z)= z^{n-18} \   B(1^-,2^-,3^-,4^-,a^+,b^+)\prod_{\substack{c=5 \\ c\neq a,b}}^n\sum_{i=1}^4\frac{[\tilde\mu_i,c]\langle i,a\rangle\langle i,b\rangle}{\langle i,c\rangle\langle c,a\rangle\langle c,b\rangle}+ {\cal O}(z^{n-19}).
\end{eqnarray}
Summing over all ways of singling out $a,b$, one recovers (\ref{n2MHV}) as desired.

\section{BCFW Deformation}\label{sec:bcfw}

The standard BCFW deformation \eqref{typicalBCFW} also fits in our framework. However, there are subtleties in the application of the multiple soft limits beyond the ones encountered in Risager-like deformations. 

Consider the BCFW deformation \eqref{typicalBCFW} applied to particles $1$ and $2$
\begin{equation}\label{BCFW}
    \{ \lambda_1,\tilde\lambda_1 \} \to \{ \lambda_1+ z \lambda_2, \tilde\lambda_1\}  \quad {\rm and}\quad  \{ \lambda_2,\tilde\lambda_2 \} \to \{ \lambda_2, \tilde\lambda_2-z\tilde\lambda_1\}.
\end{equation}
In the language of section \ref{sec:map} we have ${\cal H}=\{ 1,2 \}$ and
\be
\mu_1 = \lambda_2,~ \tilde\mu_1 = 0, ~\mu_2=0,~ \tilde\mu_2 = -\tilde\lambda_1.
\ee
Once again consider the case of graviton amplitudes, i.e. $s=2$. Moreover, let $h_1=+2$ and $h_2=-2$. The function $F(z)$ becomes
\be
F(z) = z^{-2}G(z)
\ee
with 
\be\label{GzBCFW} 
G(z) = A(\{ \lambda_2 + \lambda_1/z,\tilde\lambda_1,+2\},\{ \lambda_2,-\tilde\lambda_1+\tilde\lambda_2/z,-2 \}, \ldots ,  \{ z^{-1/2}\lambda_a,z^{-1/2}\tilde\lambda_a, h_a \},\ldots ).
\ee 
Now we are supposed to study the large $z$ behavior of $G(z)$. In this regime, particles $3,4,\ldots ,n$ become soft and particles $1$ and $2$ are in a forward limit configuration. As in previous sections, a naive application of soft theorems is not valid. Since the ``hard" particles have opposite helicities the smallest non-vanishing amplitude we can use to split the multiple soft limit is a three-particle one.
We then arrive at the formula 
\be\label{bcfwG} 
G(z) = \!\sum_{a=3}^n A(1^+,2^-,a)(z)\prod_{\substack{e=3 \\ e\neq a}}^n {\cal S}_e(z) + \ldots ,
\ee 
where once again we use Weinberg's soft factor \eqref{weinb} 
\be
{\cal S}_a = \sum_{\substack{b=1 \\ b\neq a}}^n \frac{[a,b]}{\langle a,b \rangle}\frac{\langle b,x \rangle\langle b,y \rangle}{\langle a,x \rangle\langle a,y \rangle}
\ee 
for $h_a=+2$ and the conjugate for $h_a=-2$. Introducing the $z$ dependence and separating the terms with $b=1$ and $b=2$ gives
\begin{align}\label{softFbcfw}\nonumber
    {\cal S}_a = & z\frac{[a,1](\langle 2,x \rangle+\langle 1,x \rangle/z)(\langle 2,y \rangle+\langle 1,y \rangle/z)}{(\langle a,2 \rangle+\langle a,1 \rangle/z)\langle a,x \rangle\langle a,y \rangle}- z\frac{([a,1]-[a,2]/z)}{\langle a,2 \rangle}\frac{\langle 2,x \rangle\langle 2,y \rangle}{\langle a,x \rangle\langle a,y \rangle} \\ & +\sum_{\substack{b=3 \\ b\neq a}}^n \frac{[a,b]}{\langle a,b \rangle}\frac{\langle b,x \rangle\langle b,y \rangle}{\langle a,x \rangle\langle a,y \rangle}.
\end{align} 

The next step is to compute the behavior of $A(1^+,2^-,a)(z)$. However, before carrying out the computation, let us announce the result $A(1^+,2^-,a)(z)={\cal O}(z^{0})$ in order to make an important remark.  

Looking term by term at ${\cal S}_a$ in \eqref{softFbcfw} would indicate that ${\cal S}_a(z)={\cal O}(z)$. Indeed, a term by term analysis would correspond to considering individual Feynman diagrams. If this were the behavior of $S_a(z)$ then \eqref{bcfwG} would imply that the large $z$ behavior of the amplitude $F(z)$ is ${\cal O}(z^{n-5})$. Let us now perform a simple counting to find out what the worse large $z$ behavior of individual Feynman diagrams is. It is easy to see that the worse large $z$ scaling comes from ``caterpillar" (or ``half-ladder") graphs, i.e, with only cubic vertices and particles $1$ and $2$ separated by $n-2$ vertices. These graphs have $n$ leaves, $n-2$ vertices and $n-3$ edges. The leaves corresponding to points $1$ and $2$ give $1/z^2$ powers each. Each vertex gives $z^2$ and each edge is a propagator which gives $1/z$. Combining all contributions gives ${\cal O}(z^{-2}\times z^{-2}\times z^{2(n-2)}\times z^{-(n-3)}) = {\cal O}(z^{n-5})$. This ${\cal O}(z^{n-5})$ behavior of individual Feynman diagrams is what makes the actual ${\cal O}(z^{-2})$ decay of graviton amplitudes so surprising as it implies that large cancellations must somehow happen.  

Of course, the attentive reader would have noticed that the leading order terms in the first and second terms of \eqref{softFbcfw} actually cancel. Given that these are the only terms with ${\cal O}(z)$ scaling, one finds the elusive cancellation we were looking for! In each of the $n-3$ soft factors ${\cal S}_a(z)$, the ${\cal O}(z)$ terms cancel out implying that ${\cal S}_a(z)={\cal O}(z^0)$. Having found the cancellation, we can complete the computation.  

Back to $G(z)$, the final step is to determine the large $z$ behavior of the tree-graviton amplitudes $A(1^-,2^+,a)(z)$. In this problem we do not know the helicity of the $a^{\rm th}$-particle so we have to consider the two possibilities. 

Consider the MHV sector first. Without loss of generality let us set $a=3$, 
\be\label{threeP} 
A(1^+,2^-,3^-)=\frac{\langle 2,3\rangle^6}{\langle 1,2\rangle^2\langle 1,3\rangle^2}.
\ee 
Introducing the $z$ dependence inherited from $G(z)$ in \eqref{GzBCFW}, one has that $\langle 2,3(z)\rangle^6 ={\cal O}(z^{-3})$, while   
$\langle 1(z),2\rangle^2={\cal O}(z^{-2})$ and $\langle 1(z),3(z)\rangle^2={\cal O}(z^{-1})$. Substituting the results into \eqref{threeP} gives $A(1^+,2^-,3^-)(z)={\cal O}(z^{0})$. The same analysis reveals the same behavior for the other helicity choice, i.e., $A(1^+,2^-,3^+)(z)={\cal O}(z^{0})$.

Now, the result that $G(z)={\cal O}(z^0)$, which is unexpected from individual Feynman diagrams, leads to the well-known behavior for the amplitude,
\be
F(z) = {\cal O}(z^{-2}) \quad {\rm as} \quad z\to \infty .
\ee

The natural question at this point is whether our computation is refined enough to produce the precise form of the $1/z^2$ term. Unfortunately the answer is negative given the fact that in the soft factors the leading order cancelled and subleading soft factors become important. In \cite{Cachazo:2014fwa}, Strominger and one of the authors provided the explicit form of the subleading soft graviton factor in terms of a differential operator. In \cite{Chakrabarti:2017ltl}, the subleading form of simultaneous soft limits was computed. It would be exciting to keep track of all these contributions and get an explicit form for the $1/z^2$ term. We leave this for future work.  

\section{Discussions}\label{sec:disc}

In this work we used the connection between hard and soft limits of graviton amplitudes to explain the BCL formula for the leading order of NMHV amplitudes under the Risager deformation. We extended the analysis to N$^2$MHV amplitudes and found perfect agreement with a computation based on the CSM gravity formula. 

Now, assuming that the large $z$ behavior of N$^{k-2}$MHV amplitudes under a $k$-shift has the form $z^{n-r(k)}$ where $r(k)$ is a linear function of $k$, one can find $r(k)$ by considering the $n=k+2$ amplitudes for which N$^{k-2}$MHV is equivalent to the $\overline{\rm MHV}$ sector.  A simple analysis using Hodges formula reveals that a $n=k+2$ $\overline{\rm MHV}$ amplitude goes as ${\cal O}(z^{-2k-4})$ as $z\to \infty$. This means that 
$r(k)=3k+6$. This leads to a ${\cal O}(z^{n-3k-6})$ behavior as $z\to \infty$. The attentive reader would have noticed that while the formula gives ${\cal O}(z^{n-18})$ for $k=4$ in agreement with the results of section \ref{sec:N2MHV}, it gives ${\cal O}(z^{n-15})$ for $k=3$ in disagreement with the known  ${\cal O}(z^{n-12})$. The reason for this discrepancy is that the $k=3$ deformation leads to $3$-point kinematics which is singular and therefore the behavior of the six-point $\overline{\rm MHV}$ amplitude as $z\to \infty$ is ${\cal O}(z^{-6})$ and not ${\cal O}(z^{-10})$. 

Now that we have a proposal for the large $z$ behavior, it is natural to conjecture that the explicit form of the leading order for any $k>3$ is given by the straightforward generalization of our NMHV and N$^2$MHV formulas \eqref{n2MHV}. So, we propose that it is given by
\be\label{nkMHV}
F(z) = z^{n-3(k+2)}\!\! \sum_{k+1\leq a<b\leq n}\!\!\!\!\! B(1^-,2^-,3^-,\ldots ,k^-,a^+,b^+)\prod_{\substack{c=k+1 \\ c\neq a,b}}^n\sum_{i=1}^k\frac{[\tilde\mu_i,c]\langle i,a\rangle\langle i,b\rangle}{\langle i,c\rangle\langle c,a\rangle\langle c,b\rangle}+ \ldots
\ee 
where $B(1^-,2^-,3^-,\ldots ,k^-,a^+,b^+)$ is the leading order in $1/z$ of the $\overline{\rm MHV}$ amplitude under the soft limit of particles $a$ and $b$ as defined in section \ref{sec:map}. A preliminary analysis of the N$^3$MHV case shows that the seven point amplitude's leading order goes as predicted by \eqref{nkMHV}. For $n=8$ we found a generalization of the solutions presented in section 4.1 for the Witten-RSV equations which leads to \eqref{nkMHV} by using the CSM formula.

The most pressing issue left open in this work is therefore a formal and direct proof of the formulas \eqref{nkMHV}. One possible approach is to extend the work on subleading soft theorems to the order needed to get the leading non-vanishing contribution in out computation.

One possible extension of this work is to complete the analysis of the large $z$ behavior of the BCFW deformation using soft theorems by including subleading terms to try and reproduce the explicit form of the $1/z^2$ term in gravity amplitudes. This computation could open the door for an expansion around infinity, in the sense of \cite{BCL}, for any graviton amplitude.

\section*{Acknowledgements}

The authors thank Dawit Belayneh for useful discussions. This research was supported in part by a grant from the Gluskin Sheff/Onex Freeman Dyson Chair in Theoretical Physics and by Perimeter Institute. The research of PL was supported in part by ANID/ POSTDOCTORADO BECAS CHILE/ 2022 - 74220031. Research at Perimeter Institute is supported in part by the Government of Canada through the Department of Innovation, Science and Economic Development Canada and by the Province of Ontario through the Ministry of Colleges and Universities.

\appendix

\bibliographystyle{JHEP}
\bibliography{references}

\providecommand{\href}[2]{#2}\begingroup\raggedright\begin{thebibliography}{10}

\bibitem{BCL}
D.~Belayneh, F.~Cachazo and P.~Leon, \emph{{Computing NMHV Gravity Amplitudes at Infinity}},  \href{https://arxiv.org/abs/2401.06114}{{\ttfamily 2401.06114}}.

\bibitem{Witten:2003nn}
E.~Witten, \emph{{Perturbative gauge theory as a string theory in twistor space}}, \href{https://doi.org/10.1007/s00220-004-1187-3}{\emph{Commun. Math. Phys.} {\bfseries 252} (2004) 189} [\href{https://arxiv.org/abs/hep-th/0312171}{{\ttfamily hep-th/0312171}}].

\bibitem{Elvang:2013cua}
H.~Elvang and Y.-t. Huang, \emph{{Scattering Amplitudes}},  \href{https://arxiv.org/abs/1308.1697}{{\ttfamily 1308.1697}}.

\bibitem{Cachazo:2004kj}
F.~Cachazo, P.~Svrcek and E.~Witten, \emph{{MHV vertices and tree amplitudes in gauge theory}}, \href{https://doi.org/10.1088/1126-6708/2004/09/006}{\emph{JHEP} {\bfseries 09} (2004) 006} [\href{https://arxiv.org/abs/hep-th/0403047}{{\ttfamily hep-th/0403047}}].

\bibitem{Britto:2004ap}
R.~Britto, F.~Cachazo and B.~Feng, \emph{{New recursion relations for tree amplitudes of gluons}}, \href{https://doi.org/10.1016/j.nuclphysb.2005.02.030}{\emph{Nucl. Phys.} {\bfseries B715} (2005) 499} [\href{https://arxiv.org/abs/hep-th/0412308}{{\ttfamily hep-th/0412308}}].

\bibitem{Britto:2005fq}
R.~Britto, F.~Cachazo, B.~Feng and E.~Witten, \emph{{Direct proof of tree-level recursion relation in Yang-Mills theory}}, \href{https://doi.org/10.1103/PhysRevLett.94.181602}{\emph{Phys. Rev. Lett.} {\bfseries 94} (2005) 181602} [\href{https://arxiv.org/abs/hep-th/0501052}{{\ttfamily hep-th/0501052}}].

\bibitem{Risager:2005vk}
K.~Risager, \emph{{A Direct proof of the CSW rules}}, \href{https://doi.org/10.1088/1126-6708/2005/12/003}{\emph{JHEP} {\bfseries 12} (2005) 003} [\href{https://arxiv.org/abs/hep-th/0508206}{{\ttfamily hep-th/0508206}}].

\bibitem{Benincasa:2007qj}
P.~Benincasa, C.~Boucher-Veronneau and F.~Cachazo, \emph{{Taming Tree Amplitudes In General Relativity}}, \href{https://doi.org/10.1088/1126-6708/2007/11/057}{\emph{JHEP} {\bfseries 11} (2007) 057} [\href{https://arxiv.org/abs/hep-th/0702032}{{\ttfamily hep-th/0702032}}].

\bibitem{Arkani-Hamed:2008bsc}
N.~Arkani-Hamed and J.~Kaplan, \emph{{On Tree Amplitudes in Gauge Theory and Gravity}}, \href{https://doi.org/10.1088/1126-6708/2008/04/076}{\emph{JHEP} {\bfseries 04} (2008) 076} [\href{https://arxiv.org/abs/0801.2385}{{\ttfamily 0801.2385}}].

\bibitem{Bjerrum-Bohr:2005xoa}
N.~E.~J. Bjerrum-Bohr, D.~C. Dunbar, H.~Ita, W.~B. Perkins and K.~Risager, \emph{{MHV-vertices for gravity amplitudes}}, \href{https://doi.org/10.1088/1126-6708/2006/01/009}{\emph{JHEP} {\bfseries 01} (2006) 009} [\href{https://arxiv.org/abs/hep-th/0509016}{{\ttfamily hep-th/0509016}}].

\bibitem{Bianchi:2008pu}
M.~Bianchi, H.~Elvang and D.~Z. Freedman, \emph{{Generating Tree Amplitudes in N=4 SYM and N = 8 SG}}, \href{https://doi.org/10.1088/1126-6708/2008/09/063}{\emph{JHEP} {\bfseries 09} (2008) 063} [\href{https://arxiv.org/abs/0805.0757}{{\ttfamily 0805.0757}}].

\bibitem{Conde:2012ik}
E.~Conde and S.~Rajabi, \emph{{The Twelve-Graviton Next-to-MHV Amplitude from Risager's Construction}}, \href{https://doi.org/10.1007/JHEP09(2012)120}{\emph{JHEP} {\bfseries 09} (2012) 120} [\href{https://arxiv.org/abs/1205.3500}{{\ttfamily 1205.3500}}].

\bibitem{Klose:2015xoa}
T.~Klose, T.~McLoughlin, D.~Nandan, J.~Plefka and G.~Travaglini, \emph{{Double-Soft Limits of Gluons and Gravitons}}, \href{https://doi.org/10.1007/JHEP07(2015)135}{\emph{JHEP} {\bfseries 07} (2015) 135} [\href{https://arxiv.org/abs/1504.05558}{{\ttfamily 1504.05558}}].

\bibitem{Zlotnikov:2017ahq}
M.~Zlotnikov, \emph{{Leading multi-soft limits from scattering equations}}, \href{https://doi.org/10.1007/JHEP10(2017)209}{\emph{JHEP} {\bfseries 10} (2017) 209} [\href{https://arxiv.org/abs/1708.05016}{{\ttfamily 1708.05016}}].

\bibitem{Saha:2017yqi}
A.~P. Saha, \emph{{Double soft limit of the graviton amplitude from the Cachazo-He-Yuan formalism}}, \href{https://doi.org/10.1103/PhysRevD.96.045002}{\emph{Phys. Rev. D} {\bfseries 96} (2017) 045002} [\href{https://arxiv.org/abs/1702.02350}{{\ttfamily 1702.02350}}].

\bibitem{AtulBhatkar:2018kfi}
S.~Atul~Bhatkar and B.~Sahoo, \emph{{Subleading Soft Theorem for arbitrary number of external soft photons and gravitons}}, \href{https://doi.org/10.1007/JHEP01(2019)153}{\emph{JHEP} {\bfseries 01} (2019) 153} [\href{https://arxiv.org/abs/1809.01675}{{\ttfamily 1809.01675}}].

\bibitem{Chakrabarti:2017ltl}
S.~Chakrabarti, S.~P. Kashyap, B.~Sahoo, A.~Sen and M.~Verma, \emph{{Subleading Soft Theorem for Multiple Soft Gravitons}}, \href{https://doi.org/10.1007/JHEP12(2017)150}{\emph{JHEP} {\bfseries 12} (2017) 150} [\href{https://arxiv.org/abs/1707.06803}{{\ttfamily 1707.06803}}].

\bibitem{Weinberg:1964ew}
S.~Weinberg, \emph{{Photons and Gravitons in $S$-Matrix Theory: Derivation of Charge Conservation and Equality of Gravitational and Inertial Mass}}, \href{https://doi.org/10.1103/PhysRev.135.B1049}{\emph{Phys. Rev.} {\bfseries 135} (1964) B1049}.

\bibitem{Mason:2009afn}
L.~J. Mason and D.~Skinner, \emph{{Gravity, Twistors and the MHV Formalism}}, \href{https://doi.org/10.1007/s00220-009-0972-4}{\emph{Commun. Math. Phys.} {\bfseries 294} (2010) 827} [\href{https://arxiv.org/abs/0808.3907}{{\ttfamily 0808.3907}}].

\bibitem{Cohen:2010mi}
T.~Cohen, H.~Elvang and M.~Kiermaier, \emph{{On-shell constructibility of tree amplitudes in general field theories}}, \href{https://doi.org/10.1007/JHEP04(2011)053}{\emph{JHEP} {\bfseries 04} (2011) 053} [\href{https://arxiv.org/abs/1010.0257}{{\ttfamily 1010.0257}}].

\bibitem{Cachazo:2013hca}
F.~Cachazo, S.~He and E.~Y. Yuan, \emph{{Scattering of Massless Particles in Arbitrary Dimensions}}, \href{https://doi.org/10.1103/PhysRevLett.113.171601}{\emph{Phys. Rev. Lett.} {\bfseries 113} (2014) 171601} [\href{https://arxiv.org/abs/1307.2199}{{\ttfamily 1307.2199}}].

\bibitem{Roiban:2004yf}
R.~Roiban, M.~Spradlin and A.~Volovich, \emph{{On the tree level S matrix of Yang-Mills theory}}, \href{https://doi.org/10.1103/PhysRevD.70.026009}{\emph{Phys. Rev. D} {\bfseries 70} (2004) 026009} [\href{https://arxiv.org/abs/hep-th/0403190}{{\ttfamily hep-th/0403190}}].

\bibitem{Cachazo:2012kg}
F.~Cachazo and D.~Skinner, \emph{{Gravity from Rational Curves in Twistor Space}}, \href{https://doi.org/10.1103/PhysRevLett.110.161301}{\emph{Phys. Rev. Lett.} {\bfseries 110} (2013) 161301} [\href{https://arxiv.org/abs/1207.0741}{{\ttfamily 1207.0741}}].

\bibitem{Cachazo:2012pz}
F.~Cachazo, L.~Mason and D.~Skinner, \emph{{Gravity in Twistor Space and its Grassmannian Formulation}}, \href{https://doi.org/10.3842/SIGMA.2014.051}{\emph{SIGMA} {\bfseries 10} (2014) 051} [\href{https://arxiv.org/abs/1207.4712}{{\ttfamily 1207.4712}}].

\bibitem{Cachazo:2014fwa}
F.~Cachazo and A.~Strominger, \emph{{Evidence for a New Soft Graviton Theorem}},  \href{https://arxiv.org/abs/1404.4091}{{\ttfamily 1404.4091}}.

\bibitem{Hodges:2009hk}
A.~Hodges, \emph{{Eliminating spurious poles from gauge-theoretic amplitudes}}, \href{https://doi.org/10.1007/JHEP05(2013)135}{\emph{JHEP} {\bfseries 05} (2013) 135} [\href{https://arxiv.org/abs/0905.1473}{{\ttfamily 0905.1473}}].

\end{thebibliography}\endgroup

\end{document}